\documentclass[twocolumn,preprintnumbers,amssymb]{revtex4}
\usepackage{graphicx}
\usepackage{dcolumn}
\usepackage{bm}
\begin{document} 
 
\newcommand{\sech}{{\rm sech}} 
\newcommand{\csch}{{\rm csch}} 
\newcommand{\beq}{\begin{eqnarray}}
\newcommand{\eeq}{\end{eqnarray}} 
\newcommand{\nn}{\nonumber} 
\def\ltap{\ \raise.3ex\hbox{$<$\kern-.75em\lower1ex\hbox{$\sim$}}\ } 
\def\gtap{\ \raise.3ex\hbox{$>$\kern-.75em\lower1ex\hbox{$\sim$}}\ } 
\def\CO{{\cal O}} 
\def\CL{{\cal L}} 
\def\CM{{\cal M}} 
\def\tr{{\rm\ Tr}}\ 
\def\CO{{\cal O}} 
\def\CL{{\cal L}} 
\def\CM{{\cal M}} 
\def\tr{{\rm\ Tr}} 
\newcommand{\bel}[1]{\be\label{#1}} 
\def\al{\alpha} 
\def\bt{\beta} 
\def\eps{\epsilon} 
\def\mn{{\mu\nu}} 
\newcommand{\rep}[1]{{\bf #1}} 
\def\be{\begin{equation}} 
\def\ee{\end{equation}} 
\def\bea{\begin{eqnarray}} 
\def\eea{\end{eqnarray}} 
\newcommand{\eref}[1]{(\ref{#1})} 
\newcommand{\Eref}[1]{Eq.~(\ref{#1})} 
\newcommand{\gsim}{ \mathop{}_{\textstyle \sim}^{\textstyle >} } 
\newcommand{\lsim}{ \mathop{}_{\textstyle \sim}^{\textstyle <} } 
\newcommand{\vev}[1]{ \left\langle {#1} \right\rangle } 
\newcommand{\bra}[1]{ \langle {#1} | } 
\newcommand{\ket}[1]{ | {#1} \rangle } 
\newcommand{\ev}{ {\rm eV} } 
\newcommand{\kev}{ {\rm keV} } 
\newcommand{\Mev}{ {\rm MeV} } 
\newcommand{\gev}{ {\rm GeV} } 
\newcommand{\tev}{ {\rm TeV} } 
\newcommand{\mev}{ {\rm meV} } 
\newcommand{\ma}{m^2_{\rm atm}} 
\newcommand{\ml}{m^2_{\rm LSND}} 
\newcommand{\tl}{\theta_{\rm LSND}} 
\newcommand{\ms}{m_\odot^2} 
\newcommand{\cta}{c_{\rm a}} 
\newcommand{\cts}{c_\odot} 
\newcommand{\sta}{s_{\rm a}} 
\newcommand{\sts}{s_\odot} 
\newcommand{\ctm}{c_{\rm m}} 
\newcommand{\stm}{s_{\rm m}} 
\newcommand{\mpl}{M_{Pl}}

 
\title{Gauge/Anomaly Syzygy and Generalized Brane World Models of Supersymmetry Breaking} 
 
 
\author{Ann E. Nelson} 
\author{Neal J. Weiner} 
\affiliation{Department of Physics, Box 1560, University of Washington, Seattle, WA 98195-1560, USA} 
 
 
\date{\today} 
 
\begin{abstract} 
In theories in which SUSY is broken on a brane separated from the MSSM matter fields, supersymmetry breaking is naturally mediated in a variety of ways. Absent other light fields in the theory, gravity will mediate supersymmetry breaking through the conformal anomaly. If gauge fields propagate in the extra dimension they, too, can mediate supersymmetry breaking effects. The presence of gauge fields in the bulk motivates us to consider the effects of new messenger fields with holomorphic and non-holomorphic couplings to the supersymmetry breaking sector. These can lead to contributions to the soft masses of MSSM fields which dramatically alter the features of brane world scenarios of supersymmetry breaking. In particular, they can solve the negative slepton mass squared problem of anomaly mediation and change the predictions of gaugino mediation.  
\end{abstract} 
 
 
\maketitle 
 

\vskip -0.2in 
\section{Introduction} 
\label{sec:intro} 
The standard model has been an exceptionally successful theory, 
explaining all observed phenomena except gravity, dark matter, and neutrino 
masses. Nonetheless, it is  improbable  that this is a 
complete effective theory of weak scale physics. It seems contrary to 
all reasonable expectations that the   Higgs boson, 
whose renormalized mass squared 
receives contributions proportional to the cutoff squared, 
should be so light compared to the  the Planck scale, the likely 
cutoff of field theory. 
 
 One candidate to upgrade the standard model to a more natural theory is weak-scale supersymmetry, in which every particle has a partner with opposite statistics. Because of this bose-fermi pairing, the quadratically divergent loops of the standard model are cancelled, making the light Higgs boson at least technically natural. 
 
Because supersymmetry has not been observed, if it exists, 
it must be a broken symmetry of nature. Supersymmetry breaking can be 
parameterized by a spurion field $X$, with non-vanishing F-component. We generically expect the operators 
\be 
\int d^4 \theta \> \frac{X^\dagger X F_i^\dagger F_j}{M_{Pl}^2} 
\ee 
where $F_i$ is a Minimal Supersymmetric Standard Model (MSSM) matter
field, and $i,j$ index 
flavor. In supergravity mediated theories, the contribution of these local 
operators to squark and slepton masses dominates over the contribution 
of long range gravitational interactions. Since there is no reason to 
expect that these operators will be flavor diagonal (or CP 
conserving), they generate potentially excessive additional contributions to FCNC and CP 
violating processes, such as  the $K_L$---$K_S$ mass difference, 
$\mu\longrightarrow e \gamma$ and the electron and  neutron EDMs. 
However, precision tests have shown no deviation from the standard 
model. Thus, supersymmetry breaking and mediation must be largely CP conserving and flavor blind. 
 
\subsection{Brane World Supersymmetry Breaking} 
\vskip -0.2in 
Beginning with \cite{Randall:1998uk} and more recently with \cite{Kaplan:1999ac,Chacko:1999mi}, models 
have been constructed in which the dangerous supergravity operators 
are absent at tree level due to {\em locality} in a fifth 
dimension. In these models, the MSSM matter fields are confined to 
 a three-brane in a fifth dimension. Supersymmetry breaking 
occurs on a different three brane, and, consequently, no higher dimension 
operators can exist at tree-level. Absence of  contact terms in the 
low energy theory (``sequestering'') requires there be no 
fields in the bulk lighter than the compactification scale, which is 
not  a generic feature of string theory \cite{Anisimov:2001zz}. However, as we 
live in a vacuum with broken supersymmetry and (possibly) non-zero 
cosmological constant, it seems that we already live in a non-generic 
vacuum, at least in the currently accepted conventional sense, so 
these issues may not be significant. Moreover, motivated by the 
AdS/CFT correspondence, Luty and Sundrum have recently demonstrated 
sequestering in certain four dimensional 
theories\cite{Luty:2001jh,Luty:2001zv}. Brane world and other scenarios of 
sequestered supersymmetry breaking have so many desirable features 
that we feel they should be taken seriously even without an explicit 
string realization. 
 
The mediation of supersymmetry breaking occurs by fields which exist
in the bulk. Gravity alone will mediate supersymmetry breaking via the
conformal anomaly (anomaly mediated supersymmetry breaking or
AMSB)\cite{Randall:1998uk,Giudice:1998xp}.  If gauge fields propagate
in the fifth dimension, and if supersymmetry breaking gives a large
F-term to a gauge singlet chiral superfield, then gaugino masses can
arise at tree level which are larger than the AMSB contribution. Then
gaugino mediation ({\~ g}MSB) can generate the soft masses of the scalar
matter fields through renormalization group effects.  
 
Each of these models has problematic issues of varying
severity. Anomaly mediation generically predicts negative slepton masses squared, although there are  solutions to this
problem
\cite{Pomarol:1999ie,Chacko:1999am,Katz:1999uw,Kaplan:2000jz,Arkani-Hamed:2000xj,EAM}.
Gaugino mediation has the stau as the lightest superpartner (LSP) over much of its parameter space \cite{Ellis:2001kg}. In all of these models, however, the setup is minimal, in particular, there are no new fields which are charged under the MSSM gauge group.  
 
In \cite{Nomura:2001ub}, it was noted that the brane world scenario can also serve as a setting for gauge mediation in which the gravitino is not the LSP. Here, the point is evident that it is natural to have messenger fields living on the brane with contact interactions with the supersymmetry breaking sector. However, phenomenologically, this model required a very small coupling between messenger fields and the SUSY breaking spurion. 
 
In what follows we will adopt the brane world setup, and we will assume the following features of the theory: 
\begin{itemize} 
\item{As in most brane world SUSY breaking scenarios, we will assume that MSSM matter fields live at one orbifold fixed point in a compact space with volume $O(10\sim 100)$}, 
\item{We will assume that gauge fields propagate in the extra
    dimension(s), although the gauginos need not necessarily get masses by coupling to a singlet as in gaugino mediation}, 
\item{As gauge fields propagate in the bulk, it is natural to expect 
    additional fields, charged under the SM, to exist either localized 
    on the brane where SUSY is broken, or in the bulk}. 
\end{itemize} 
 
\vskip -0.2in 
\section{Gauge/Anomaly Mediated Syzygy} 
\vskip -0.1in
We will begin by studying the simplest case, in which there are no
singlets with F terms. The anomaly mediated
contribution to the slepton mass squared is negative. However, we now have
messenger fields which can alter the slepton mass prediction.

Let us consider matter fields $M$ and $\overline M$, transforming nontrivially under the  MSSM gauge group, localized on the
supersymmetry beaking brane. Below the compactification scale, we have  the ordinary supergravity operators 
\be 
\int d^4 \theta \> \frac{\lambda X^\dagger X M^\dagger M}{M_*^2}+\frac{\overline \lambda X^\dagger X \overline M^\dagger \overline M}{M_*^2}, 
\ee which  generate soft masses for the messenger fields $\delta m^2 =
\lambda M_*^n V_n m_{3/2}^2$. These terms do not violate the R-symmetry of the theory and will not contribute to the gaugino masses. However, they will contribute to the soft masses of the MSSM matter fields by an amount \cite{Poppitz:1997xw,Arkani-Hamed:1998kj} 
\be 
m^2_i = - \sum_a\frac{g_a^4}{128 \pi^4} S_Q C_{ai} {\rm Str} M_{mess}^2 \log \left(\frac{\Lambda^2 _{UV}}{m^2_{IR}}\right), 
\ee 
where $m^2_{IR}$ is the mass of the messenger fields, $S_Q$ is the
Dynkin index of the messenger representation, and $C_{ai}$ is the
quadratic Casimir for the representation of the MSSM matter field in
question. When the log terms are large, these terms should be resummed
to yield
\be \sum_a\frac{S_Q C_{ai} {\rm Str} M_{mess}^2 \left(
    g_a^2(m^2_{IR})-g_a^2(\Lambda^2 _{UV})\right)}{8 \pi^2 b_a} \ ,
\ee
 where $b_a$ is the coefficient of the one-loop $\beta$ function for $g_a$ above the messenger scale.

When the messengers are localized on the SUSY breaking brane the
contribution to the supertrace is naturally $ \sim \lambda F^2/M_*^2 =
\lambda  F^2 M_*^n V_n/M_{Pl}^2 =  \lambda m_{3/2}^2 M_*^n V_n$. Since
the anomaly mediated contribution also occurs at two loops,  it will
 be smaller by the volume factor, which can be $O(10\sim
100)$. Moreover, the messenger contribution can receive a logarithmic enhancement
relative to the anomaly mediated contributions. The sign of the
messenger contribution depends  on the sign of the nonrenormalizable operator, but can be positive. 
 
To address the negative slepton mass squared problem,
while retaining the phenomenologically desirable feature that gaugino
and scalar masses are about the same size, these effects should be
comparable. One can simply assume that the coefficients of the
messenger/SUSY breaking contact operators are small to compensate for the volume factor, but a more natural solution would be to assume that the messenger fields themselves propagate in the extra dimension, in which case the coefficients $\lambda, \overline \lambda$ are suppressed by precisely this same volume factor. The alignment or ``syzygy'' of these contributions appears fortuitous, because they have distinct origins. Yet within  this model, it occurs completely naturally. 
However, as pointed out in ref. \cite{Chacko:2001jt}, in this case contact terms between messengers and MSSM fields could lead to flavor violation and FCNC at one loop, unless suppressed by a mechanism for repelling the messengers from our brane. Several such mechanisms were suggested in ref. \cite{Chacko:2001jt}.
 
\vskip -0.2in 
\section{Messengers in Gaugino Mediation} 
\vskip -0.1in
If  singlets with F terms exist on the SUSY breaking brane,
then gauge particles naturally
have contact terms with the singlet which give rise to supersymmetry
breaking gaugino masses
which dominate the anomaly mediated contributions
\cite{Kaplan:1999ac,Chacko:1999mi}.  We are now also allowing messenger fields. What effects can we expect? 
 
The first important point is that the gaugino mass, being localized on 
a brane, is volume suppressed \cite{footnote1}. 
 If the messengers are localized on the brane (as opposed to 
propagating in the bulk, as discussed in the 
previous section), then their contributions to scalar masses are naturally larger than the 
gaugino mediated terms. 
 
Let us assume that the messenger fields have masses at some lower scale, $m$. Then the nonrenormalizable operators already described give rise to contributions modifying the sfermion masses. In terms of the gravitino mass and the size of the extra dimensional volume 
\be 
m^2_i \sim - \sum_a\frac{g_a^4}{128 \pi^4} S_Q C_{ai} m_{3/2}^2 V_n M_*^n \log
\left( \frac{\Lambda^2 _{UV}}{m^2_{IR}}\right) \ . 
\ee
In contrast, the gaugino mass is given by  
\be 
m_{\tilde g} \sim \frac{m_{3/2}}{\sqrt{V_n M_*^n}} 
\ee 
The usual one-loop RG contribution to soft masses is  
\be 
\delta m^2 \sim \frac{g^2}{16 \pi^2} \frac{m_{3/2}^2}{V_n M_*^n} \log (\frac{\Lambda^2}{m_{IR}^2}) 
\ee 
The relative strength of these effects is then (taking the logarithms to be of comparable size) 
\be 
\frac{messenger \> contributions}{{\tilde g}MSB} \sim \frac{g^2}{16 \pi^2} S_Q (V_n M_*^n)^2. 
\ee 
For $V_n M_*^n \sim 10$, these can be of comparable size, dramatically 
changing the spectrum of gaugino mediation. In particular, if these 
effects are positive, the scalar masses can be naturally heavier than 
the gaugino masses, leading to, e.g., a bino LSP over broader ranges 
of parameter space. The bino is preferable to the stau as LSP  since 
 stable charged relics are extremely constrained \cite{Kudo:2001ie}, while a stable bino is an 
acceptable dark matter candidate.  
 
Up to this point, we have only considered the effects of 
non-holomorphic masses for the messengers. With messengers coupled to a singlet, we have the expectation of additional holomorphic masses as well. 
 
In particular, we must assume that the messengers are sufficiently
light compared with the fundamental scale of the theory so that we can
treat them within effective field theory. We will not explain the origin of their mass here, but simply introduce a superpotential operator, 
\be 
W \supset m M \overline M 
\ee 
Given the presence of the singlet field $X$, we  expect the presence of the additional operator 
\be 
W \supset \frac{X}{M_*} m M \overline M 
\ee 
If the gaugino mass arose from an order one coupling to $X$, this
would lead to insignificant contributions to soft masses. However the gaugino mass is volume suppressed, and $F/M \sim m_{3/2} \sqrt{V_n M^n_*}$. 
These terms will generate contributions to the scalar and gaugino
masses which are of the conventional gauge mediated sort \cite{Dine:1993yw,Dine:1995vc,Dine:1996ag}.
   
The additional gaugino masses scale as 
\be 
\delta m_{\tilde g_i}=\frac{\alpha_i S_Q}{4 \pi} \frac{F}{M}\simeq \frac{\alpha_i}{4 \pi}S_Q  m_{3/2} \sqrt{V_n M_*^n} 
\ee 
which compares with the tree level piece 
\be 
\frac{messenger \> contribution}{tree \> level} \sim \frac{\alpha_i}{4 \pi} S_Q V_n M_*^n, 
\ee 
so that these effects can be competitive. Of course, if all the new pieces are simply proportional to gauge couplings, these effects are consistent with a unified tree level piece at some higher scale. The distinctions between this case and traditional gaugino mediation come from the additional contributions to the scalar masses from the holomorphic terms and from eq. (7), both of which are volume enhanced. 
 
\subsection{$\mu$ and $B \mu$} 
\vskip -0.2in
Brane world models such as gaugino mediation in which the gravitino mass is comparable to the weak scale have a natural solution to the $\mu$ problem. If Higgs fields propagate in the bulk, one can simply employ the Giudice-Masiero mechanism \cite{Giudice:1988yz} by coupling the Higgs fields directly to the supersymmetry breaking fields \cite{Chacko:1999am}. However, even when the Higgs field is localized to a brane isolated from the supersymmetry breaking, it is trivial \cite{Nomura:2001ub}. One simply includes a term in the K\" ahler potential 
\be 
\int d^4 \theta \> \phi^\dagger \phi H_u H_d, 
\ee 
where $\phi$ is the conformal compensator. Because $\vev{\phi} = 1+ \theta^2 m_{3/2}$ this generates both a $\mu$ and $B \mu$ term. However, in the limit of gauge-anomaly syzygy, in which case the soft masses are down by a loop factor from the gravitino mass, this is unworkable \cite{Randall:1998uk}. 

\vskip -0.2in 
\section{Parameter space and phenomenology} 
\vskip -0.1in
\label{sec:pheno} 
A complete treatment of the phenomenology of these models will be left
for future work, but we will briefly comment on the most significant
qualititative effects. There are  two main limiting cases. In the
first case, which we refer to as ``gaugino-like'', the anomaly
mediated contributions are relatively small when compared with
contributions arising from singlets. In the second, ``anomaly-like''
case, the gravitino is heavier than the weak scale by a loop factor,
and holomorphic mass terms are small.
 
A great advantage of the  anomaly-like case over  minimal
anomaly mediation is that sleptons are not tachyonic.
Still, the model is highly predictive for the soft masses of the
sfermions and gauginos, and retains a solution to the SUSY flavor problem. However, as we have not yet included a solution to the $\mu$ problem, we cannot speak reliably of the soft mass of the Higgs fields. 
 
The gaugino-like case offers simple solutions to the $\mu$ problem. Unlike traditional gaugino mediation, the gauge mediated (both holomorphic and non-holomorphic) contributions can compete with the gaugino mediated contributions, changing the size of the sfermion masses relative to the gaugino masses, while maintaining flavor blindness. 
As in the original gaugino mediated framework, the Higgses can either
be on our brane or in the bulk. In the latter case the operators 
\be 
\int d^4 \theta \> \frac{X}{\mpl} (\xi_u H_u^\dagger H_u + \xi_d H_d^\dagger H_d)  
\ee 
give  A-terms proportional to the Yukawas and a $B\mu$
parameter. Furthermore, contact terms with the SUSY breaking brane can
contribute to
the soft Higgs masses.

We can include all cases as limits of the  following parameter space,
although in most scenarios only a few parameters will be relevant: We
have $\mu$, $B \mu$, $m_{3/2}$, $m_{1/2}$,  $\lambda$ (the strength of
non-holomorphic operators),   $S_Q$, $F/M$  $M$, $\Lambda$, soft Higgs
mass squared parameters $m_{H_d}^2$, $m_{H_u}^2$, $\xi_u$,
$\xi_d$. After fixing $m_Z$, we have twelve continuous parameters
and one discrete parameter (the sign of $\mu$).

The most predictive scenario is the gaugino-like case with Higgs
confined to our brane. All soft SUSY breaking parameters can then be
computed from $\mu$,  $m_{3/2}$ (or equivalently $B$), $m_{1/2}$,
$\lambda$, $\Lambda$, $S_Q$, $F/M$ and $M$. One combination can be fixed by
$m_Z$, leaving seven continuous and one discrete parameter.  

Another  predictive case is when there are no holomorphic mass
terms (anomaly-like). In this case, however we do not have a preferred
solution to the $\mu$ problem, so we leave $m_{H_d}^2$, $m_{H_u}^2$,
and $B \mu$ free. The other parameters are $\mu$, $m_{3/2}$,
$\lambda$, $\Lambda$, and $M$. After fixing $m_Z$ we are left with
seven continuous and one discrete parameter. 

One advantage of including messengers is that there are now
significant contibutions to slepton masses arising at short distance
where their gauge couplings are as large as those of the
squarks. Thus, unlike  traditional anomaly or gaugino mediation, which
both predict relatively light sleptons, slepton and squark masses may
now be comparable. Indeed, if these contributions 
arise above the GUT scale, they might be universal.
Such a scenario would  require less
fine-tuning than either AMSB or { \~ g}MSB. 

 While the number of
parameters is slightly larger than in mSUGRA, generalized brane world
models have a much richer
phenomenology,  a basis in a well defined framework, and no
 supersymmetric flavor problem. In addition, if CP is only broken on
 our brane, there are no phases in the soft parameters and the SUSY CP problem is solved.
 
The inclusion of AMSB and {\~g}MSB in the same framework --- both as realistic models --- with a limit that looks like a  version of gauge mediation,  is a  step towards the ``generalized model space'' previously proposed \cite{dek}. 

\vskip -0.2in 
\section{Conclusions} 
\vskip -0.1in 
In brane world models of supersymmetry breaking where gauge fields propagate in the bulk, it is natural to consider the presence of additional messenger fields transforming under the MSSM gauge group. The effects of these fields can drastically change the features of well known brane world models. In particular, it can solve the negative slepton mass squared problem within anomaly mediated supersymmetry breaking, and change the mass relationships of gaugino mediation. 
 
The phenomenology of these models depends strongly on the inputs: the size of the extra dimensions and the mass scale of the messengers in particular. A study of the phenomenology of these models would be  useful. 
\par 
\vskip 0.1in \par
 
{\bf \noindent Acknowledgements}  
 
\vskip 0.05in 
 
\noindent While this manuscript was being prepared ref. \cite{Chacko:2001jt} appeared, with overlapping results. This work was partially supported by the DOE under contract DE-FGO3-96-ER40956.  We thank Y. Nomura, Z. Chacko and P. Fox for useful conversations.
 
\bibliography{syzygy} 

\begin{thebibliography}{25}
\bibitem[{foo()}]{footnote1}
\bibinfo{journal}{Equivalently, we could work with naive dimensional analysis,
  in which the volume suppression appears in the four dimensional gauge
  coupling. However, in such a setup, the brane localized term would have been
  small from the outset.}
\end{thebibliography}
 
\bibliographystyle{apsrev} 
 
\end{document}